\documentclass[fleqn,12pt,twoside]{article}
\usepackage{espcrc1}

\usepackage{graphicx}
\usepackage{wrapfig}
\usepackage[figuresright]{rotating}

\newcommand{\AmS}{{\protect\the\textfont2
  A\kern-.1667em\lower.5ex\hbox{M}\kern-.125emS}}

\title{Charge Fluctuations at Mid-Rapidity in Au+Au Collisions in the PHENIX Experiment at RHIC}

\author{
J. Nystrand\address[MCSD]{Div. of Cosmic and Subatomic Physics, Department of Physics, Lund University, 
SE-22100 Lund, Sweden} 
for the PHENIX Collaboration\thanks{for the full PHENIX Collaboration author list and acknowledgements, 
see Appendix "Collaborations" of this volume.}
}

\begin{document}

% typeset front matter
\maketitle

\begin{abstract}
Charged particle event-by-event fluctuations have long been considered as a possible 
signature of quark gluon plasma formation in ultra-relativistic heavy-ion collisions \cite{Gyulassy}. 
That the interest in fluctuations remains high in the era of RHIC is evidenced by the 
large number of recent theoretical \cite{recent_th,heiselberg,Zhang}, as well as 
experimental \cite{ppg007,ppg005,NA49STAR}, studies published since the previous Quark Matter conference. 
Here, new results from the PHENIX experiment on event-by-event fluctuations in net charge and $<p_T>$ in 
Au+Au interactions at $\sqrt{s_{NN}} =$~200 GeV will be presented. 
\end{abstract}

\section{Net charge fluctuations}

The idea to study net charge fluctuations as a signal for quark gluon plasma 
formation was proposed about two years ago \cite{AHM_JK}. It was predicted 
that the fluctuations in the net charge in a local region of phase space should be drastically reduced 
if a plasma was formed in the collisions. The arguments are based on the fact that the electric charge 
is more evenly distributed in a plasma, and it is predicted that this distribution should survive the 
transition back to the hadronic state, if a sufficiently large region of phase space ($\Delta y \sim$~1)
is considered. 

Net charge fluctuations in Au+Au interactions at $\sqrt{s_{NN}} =$~130~GeV have previously been studied 
by PHENIX \cite{ppg007}. The same detectors and experimental techniques are 
used in the present analysis for Au+Au at $\sqrt{s_{NN}} =$~200~GeV, with the exception that tracks are 
required to have an associated hit in the outermost layer of pad chambers ($\sim$5~m from the 
interaction point), resulting in improved track quality at the expense of a slightly reduced acceptance. 
The outermost layer of pad chambers was not installed in the west tracking arm in the run at 130~GeV.

For a given centrality selection, as determined by the PHENIX beam-beam counters (BBC) and zero-degree 
calorimeters (ZDC), one reconstructs $n_+$ positive and $n_-$ negative particles in an individual event. 
The fluctuations are studied using the variable $v(Q)$, which is calculated as the ratio of the variance 
of the net charge, $Q = n_+ - n_-$, to the mean of the total charged multiplicity, $n_{ch} = n_+ + n_-$, 
\begin{equation}
v(Q) = \frac{V(Q)}{<n_{ch}>} \; .
\end{equation}

\begin{wrapfigure}{l}{0.55\textwidth}
\begin{center}
\vspace{-1.1cm}
\includegraphics[width=0.55\textwidth]{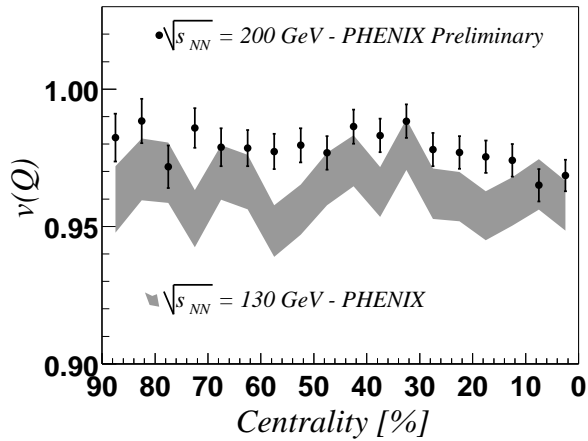}
\caption{$v(Q)$ vs. centrality for data at $\sqrt{s_{NN}} =$~130 and 200 GeV. The rightmost bin 
corresponds to the 0-5\% most central collisions. The band (the width represents the statistical error) 
shows the result from 130 GeV \cite{ppg007}. The results at 200 GeV are for $| \eta | <$~0.35, 
0.3~$< p_T <$~2.0~GeV/c, $\Delta \varphi = \pi/2$.}
\label{fig:centrality}
\end{center}
\end{wrapfigure}
For random 
emission of particles, $v(Q) = 1$. In a system with a random distribution of particles in phase space and with 
global charge conservation, one expects $v(Q) = (1-p)$ if a fraction $p$ of all produced particles are detected. 
The experimentally observed values of $v(Q)$ for different centralities are shown in Figure~1. 

One notes that $v(Q)$ shows only a small deviation from the value 1 for all centrality selections. 
Data at 130 GeV show no centrality dependence within the statistical errors while at 200 GeV 
a slight decrease in $v(Q)$ with increasing centrality cannot be excluded. The value of $v(Q)$ is far 
from what has been predicted for a quark gluon plasma ($v(Q) \sim$~0.2) and a hadron gas 
($v(Q) \sim$~0.7) \cite{AHM_JK}. It should be noted that these predictions are for a rapidity 
interval $\Delta y \sim$~1 with full azimuthal acceptance, whereas PHENIX covers $\Delta \eta =$~0.7  
in pseudo-rapidity and $\Delta \varphi = \pi/2$ in azimuth. 

The question is therefore if and to what extent the measured values of $v(Q)$ reflect only global charge 
conservation or if any other contributions, e.g. from resonance decays in a hadron gas, are important. The reduction 
in $v(Q)$ is expected to scale with the acceptance. This is the case if only global charge conservation is 
considered, as was discussed above. It is also true for a reduction from resonances in a hadron gas. 
The reason is that a certain geometrical coverage is needed to detect all charged particles from the 
decay of a neutral resonance. The acceptance scaling of $v(Q)$ for a quark gluon plasma is not known, 
hence a theoretical model would be desirable to be able to do experimental comparisons. 

\begin{figure}[htb]
\vspace{-0.4cm}
\begin{minipage}[t]{75mm}
\centerline{\includegraphics[height=55mm]{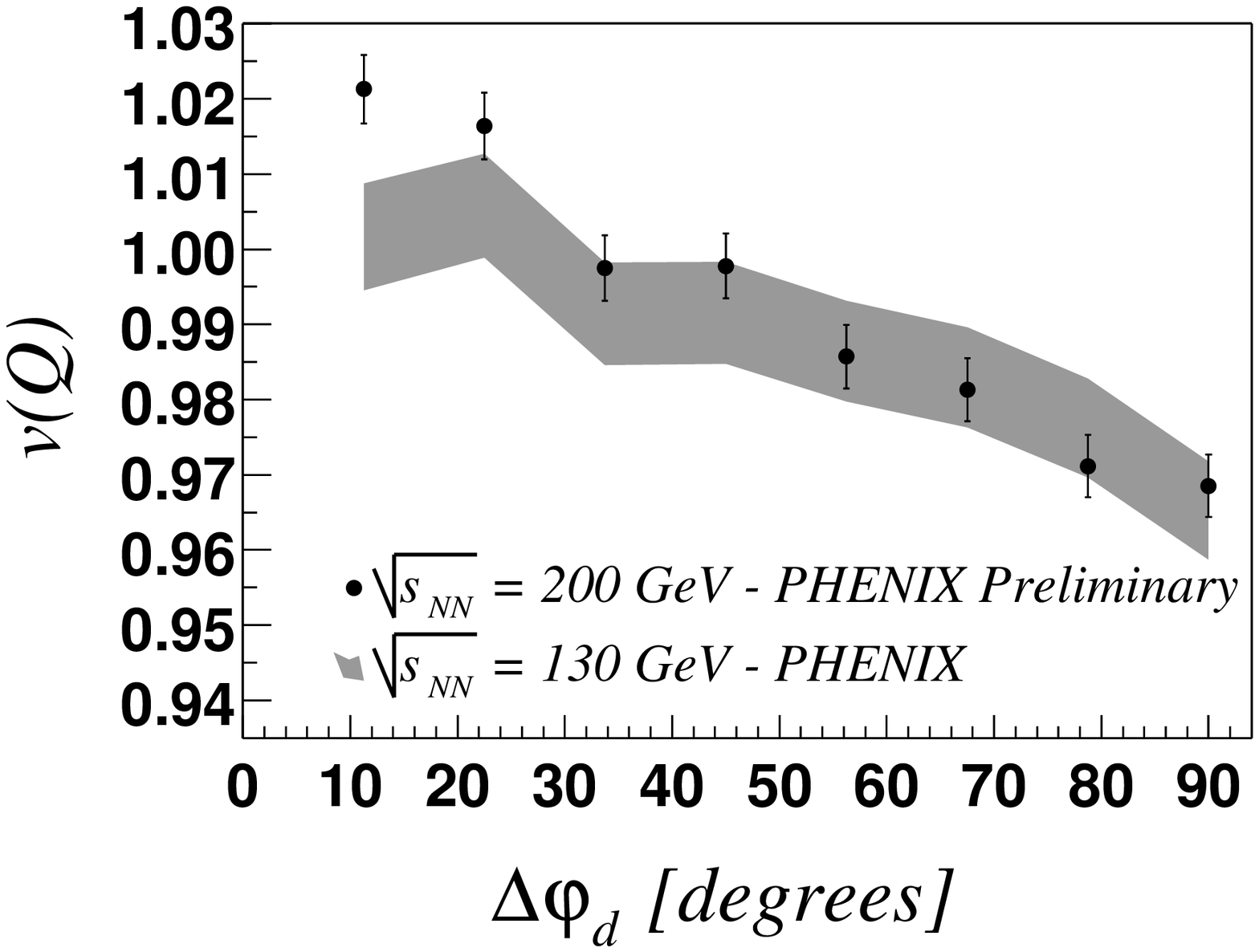}}
\end{minipage}
\hspace{\fill}
\begin{minipage}[t]{75mm}
\centerline{\includegraphics[height=55mm]{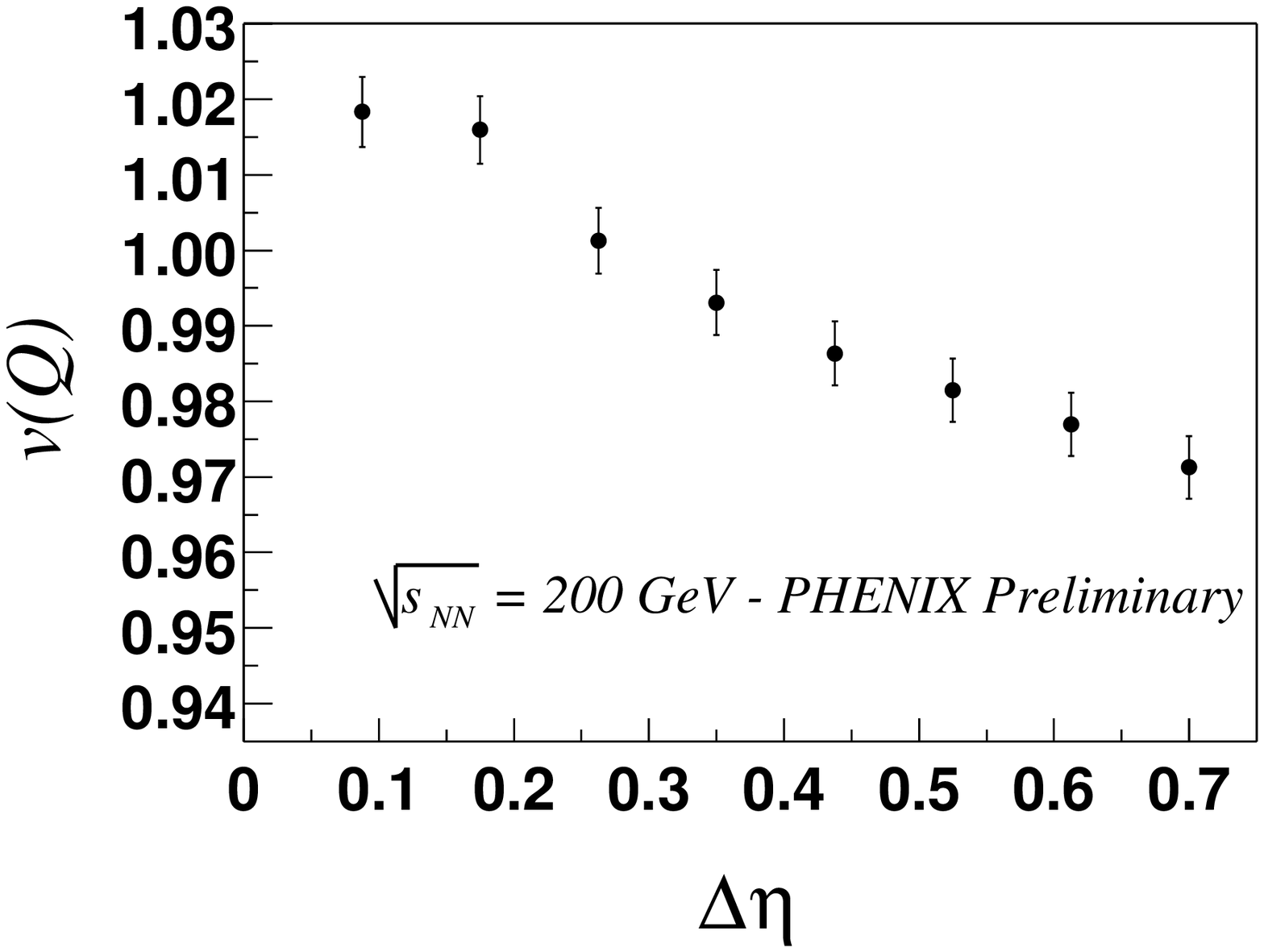}}
\end{minipage}
\caption{$v(Q)$ vs. $\Delta \varphi_d$ (left) and vs. $\Delta \eta$ (right) for the  10\% most central collisions.}
\end{figure}

Experimentally, it is found that $v(Q)$ does scale with the acceptance, as is illustrated in Figure 2, where $v(Q)$ 
is plotted versus varying acceptance windows in azimuth and pseudo-rapidity. The trend is very similar in both cases. 
The variable $\Delta \varphi_d$ corresponds to the azimuthal coverage that is used; it ranges from 0$^0$ to 
90$^0$ in one central arm of the PHENIX spectrometer. It is not currently understood why $v(Q)$ at 
$\sqrt{s_{NN}} =$~200~GeV exceeds one for small acceptances. 

For data at 130 GeV it was estimated that global charge conservation alone should lead to a reduction in $v(Q)$ of 
about 1\%. The observed value (0.965$\pm$0.007), which was found to be in good agreement with RQMD \cite{Sorge}, 
was more than 3 standard deviations lower than the charge conservation estimate, thus indicating that charged tracks 
emanating from decays of neutral resonances are necessary to fully understand the measurements. No additional 
suppression to what was 
predicted by RQMD was observed. The reduction due to global charge conservation is approximately the same at 
$\sqrt{s_{NN}} =$~200 and 130 GeV, which is also true for the fluctuations in RQMD \cite{Zhang}. The preliminary 
conclusion is that both at $\sqrt{s_{NN}} =$~130 and 200 GeV we do not observe any anomalous suppression in $v(Q)$ 
that cannot be explained by hadronic models.

\section{Fluctuations in $<p_T>$}

Fluctuations associated with a QCD phase transition could lead to anomalous event-by-event fluctuations in the 
average transverse momentum \cite{heiselberg}. Fluctuations in $<p_T>$ were studied by PHENIX at 130 GeV but no excess 
above the random contribution was observed \cite{ppg005}. The present analysis uses the same techniques but 
with increased statistics, improved track quality cuts and better background rejection.

The observed fluctuations are compared with the expectation from statistically independent particle 
emission through the use of mixed events, as described in \cite{ppg005}. The quantity  
\begin{equation}
\omega = \frac{ \sqrt{ <p_T^2> - <p_T>^2 } }{ <p_T> } 
\end{equation}
is calculated for real and mixed events. The deviations from stochastical fluctuations are quantified through 
the measure $F_{T}$:
\begin{equation}
F_{T} = \frac{ \omega_{data} - \omega_{random} }{ \omega_{random} } \; .
\end{equation}
A non-zero value of $F_{T}$ indicates non-statistical fluctuations.

$F_{T}$ is plotted as a function of centrality in the left part of Figure 3. At $\sqrt{s_{NN}} =$~200~GeV a 
statistically significant signal is observed which was not the case at 130~GeV. The value 
of $F_{T}$ has a maximum in semi-central collisions.  

\begin{figure}[htb]
\vspace{-0.35cm}
\begin{minipage}[t]{75mm}
\includegraphics[height=50mm]{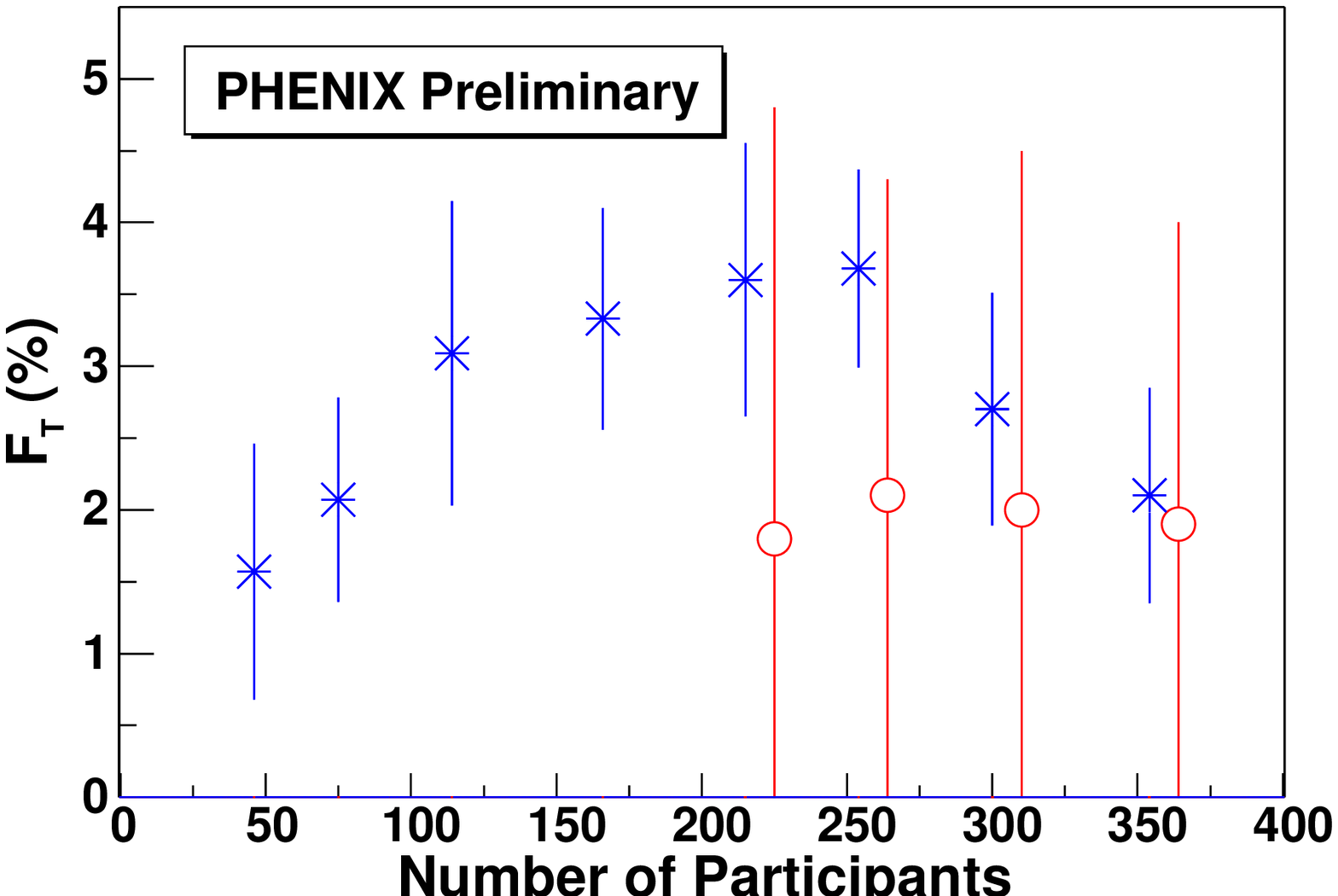}
\label{fig:largenenough}
\end{minipage}
\hspace{\fill}
\begin{minipage}[t]{75mm}
\includegraphics[height=50mm]{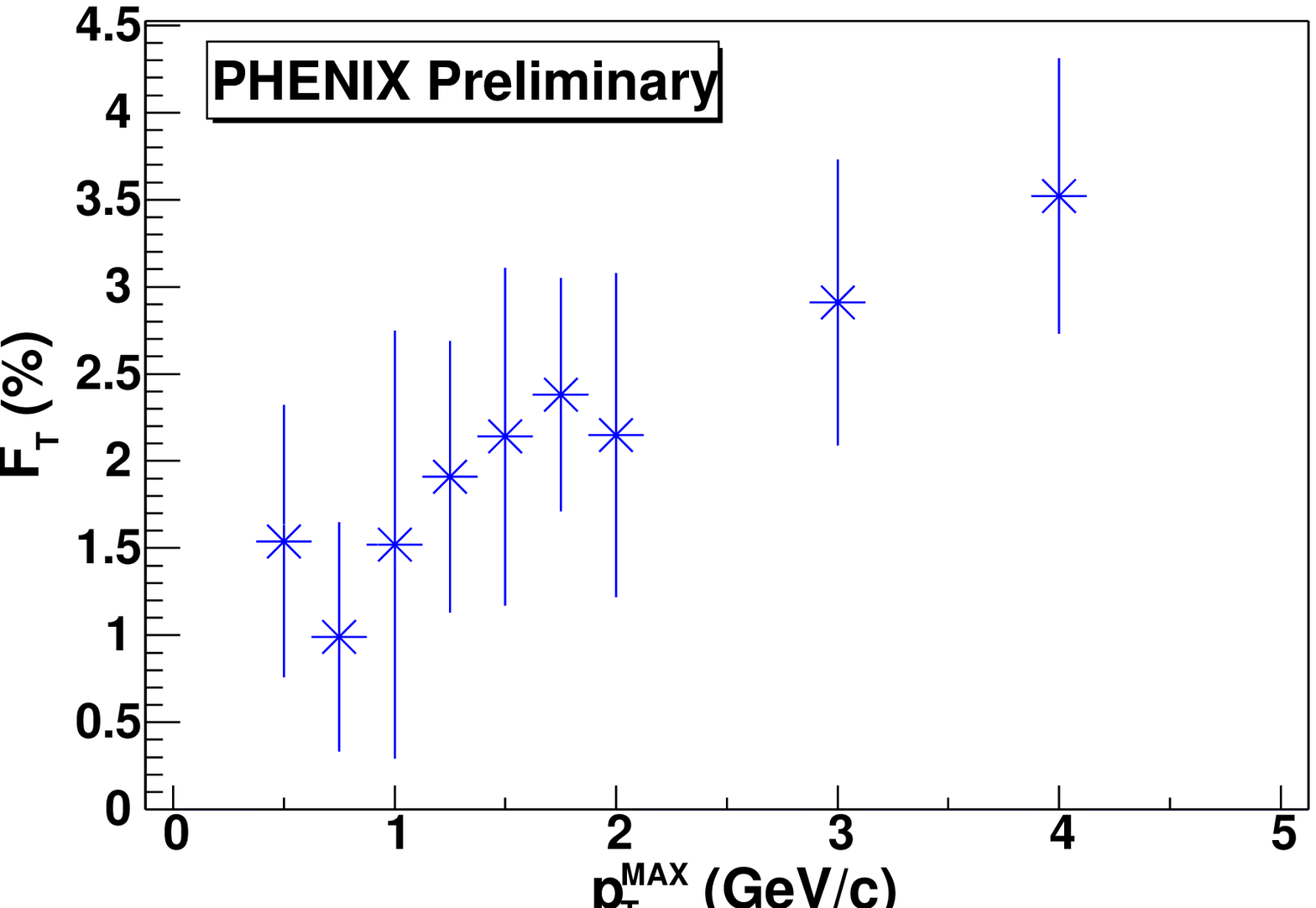}
\label{fig:toosmall}
\end{minipage}
\caption{$F_T$ vs. centrality (left) and $F_T$ vs. $p_T^{MAX}$ for central collisions (right). The asterisks 
and open circles are for Au+Au at $\sqrt{s_{NN}} =$~200 and 130~GeV, respectively.}
\end{figure}

The right part of Figure 3 shows the variation of $F_{T}$ with the $p_T$-range used to calculate $<p_T>$, 
from 0.2~GeV/c to $p_T^{MAX}$, plotted as function of $p_T^{MAX}$. As can be seen, the 
value of $F_{T}$ increases as the $p_T$ range is extended to higher values. 

The variation of $F_{T}$ with centrality and $p_T$ is similar to what one would expect for a process connected  
with elliptic flow. A simulation, based  on the observed values of $v_2$ \cite{PHENIXFlow}, of the contribution from 
elliptic flow to $F_{T}$ was therefore performed. The result indicates 
that elliptic flow is not responsible for the observed $F_{T}$ signal.

\section{Conclusions}

The analysis of net charge fluctuations at $\sqrt{s_{NN}} =$~200 GeV has yielded results similar to those at 130 GeV. 
No additional suppression of the fluctuations beyond what is predicted by hadronic models is observed. 
Fluctuations in the mean transverse momentum, as measured through the quantity $F_{T}$, show a positive  
signal at $\sqrt{s_{NN}} =$~200 GeV, peaking in semi-central collisions.


\begin{thebibliography}{9}

\bibitem{Gyulassy} M. Gyulassy, Nucl. Phys. A 400 (1983) 31c.

\bibitem{recent_th} H.~Heiselberg and A.~Jackson, Phys. Rev. C 63 (2001) 064904; 
F.W.~Bopp and J.~Ranft, Eur. Phys. J. C 22 (2001) 171-177, 
Acta Phys. Polon. B 33 (2002) 1505-5120; B.~M{\"u}ller, Nucl. Phys. A 702 (2002) 281c-290c; 
V.~Koch, M.~Bleicher, and S.~Jeon, Nucl. Phys. A 698 (2002) 261c-268c, Nucl. Phys. A 702 (2002) 291c-298c;
M.~Asakawa, U.~Heinz, and B.~M{\"u}ller, Nucl. Phys. A 698 (2002) 519c-522c.

\bibitem{heiselberg} H. Heiselberg, Phys. Rep. 351 (2001) 161.

\bibitem{Zhang} Q.H.~Zhang, V.~Topor~Pop, S.~Jeon, and C.~Gale, Phys. Rev. C 66 (2002) 014909.

\bibitem{ppg007} K. Adcox et al. PHENIX Collaboration, Phys. Rev. Lett. 89 (2002) 082301.

\bibitem{ppg005} K. Adcox et al. PHENIX Collaboration, Phys. Rev. C 66 (2002) 024901.

\bibitem{NA49STAR} S.V. Afanasiev et al. NA49 Collaboration, Nucl. Phys. A 698 (2002) 104c;
J.G. Reid et al. STAR Collaboration, Nucl. Phys. A 698 (2002) 611c. 

\bibitem{AHM_JK} M. Asakawa, U. Heinz, and B. M{\"u}ller, Phys. Rev. Lett. 85 (2000) 2072; 
S. Jeon and V. Koch, Phys. Rev. Lett. 85 (2000) 2076. 

\bibitem{Sorge} H. Sorge, Phys. Rev. C 52 (1995) 3291. 

\bibitem{PHENIXFlow} S. Esumi for the PHENIX Collaboration, these proceedings. 

\end{thebibliography}
\end{document}